\documentclass[aps,prx,twocolumn,preprintnumbers,amsmath,amssymb,superscriptaddress,showpacs]{revtex4-2}

\usepackage{graphicx}
\usepackage{epstopdf}
\usepackage{amsmath}
\usepackage{multirow}
\usepackage{xcolor}
\usepackage{ulem}
\usepackage[version=4]{mhchem}
\usepackage{hyperref}
\hypersetup{
    colorlinks=true, 
    linkcolor=blue,  
    citecolor=blue,   
    filecolor=magenta, 
    urlcolor=blue,   
}

\begin{document}

\title{Screening phonon-mediated superconductors from static orbital Hamiltonians}

\author{Jian-Feng Zhang}\email{jianfeng.zhang@hpstar.ac.cn}
\affiliation{ Center for High Pressure Science and Technology Advanced Research, Beijing 100193, China. }

\author{Ze-Feng Gao}
\affiliation{ School of Physics and Beijing Key Laboratory of Opto-electronic Functional Materials \& Micro-nano Devices, Renmin University of China, Beijing 100872, China }
\affiliation{ Key Laboratory of Quantum State Construction and Manipulation (Ministry of Education), Renmin University of China, Beijing 100872, China }

\author{Xiao-Qi Han}
\affiliation{ School of Physics and Beijing Key Laboratory of Opto-electronic Functional Materials \& Micro-nano Devices, Renmin University of China, Beijing 100872, China }
\affiliation{ Key Laboratory of Quantum State Construction and Manipulation (Ministry of Education), Renmin University of China, Beijing 100872, China }

\author{Dingshun Lv}
\affiliation{ Field Quantum Research Institute, Beijing 100084, China }

\author{Miao Gao}
\affiliation{Department of Physics, School of Physical Science and Technology, Ningbo University, Zhejiang 315211, China}

\author{Kai Liu}
\affiliation{ School of Physics and Beijing Key Laboratory of Opto-electronic Functional Materials \& Micro-nano Devices, Renmin University of China, Beijing 100872, China }
\affiliation{ Key Laboratory of Quantum State Construction and Manipulation (Ministry of Education), Renmin University of China, Beijing 100872, China }

\author{Xinguo Ren}
\affiliation{ Institute of Physics, Chinese Academy of Sciences, Beijing 100190, China}

\author{Zhong-Yi Lu}\email{zlu@ruc.edu.cn} 
\affiliation{ School of Physics and Beijing Key Laboratory of Opto-electronic Functional Materials \& Micro-nano Devices, Renmin University of China, Beijing 100872, China }
\affiliation{ Key Laboratory of Quantum State Construction and Manipulation (Ministry of Education), Renmin University of China, Beijing 100872, China }

\author{Tao Xiang}\email{txiang@iphy.ac.cn}
\affiliation{ Institute of Physics, Chinese Academy of Sciences, Beijing 100190, China}
\affiliation{ School of Physical Sciences, University of Chinese Academy of Sciences, Beijing 100049, China}

\date{\today}


\begin{abstract}
 The first-principles search for superconductors is severely limited by the high cost of electron-phonon coupling (EPC) calculations. Here we develop a low-cost, physically transparent framework that identifies strong-EPC materials directly from static orbital-based Hamiltonians without explicit phonon perturbation calculations. Verification using density functional perturbation theory (DFPT) for representative superconductors shows that the framework captures semi-quantitatively the EPC scale at substantially lower computational cost. Applied to more than 36,000 compounds in the MattKeyBond database, it identifies 34 dynamically stable superconducting candidates with calculated $T_c > 10$ K after DFPT verification. These candidates reveal two distinct routes to relatively high-$T_c$ superconductivity: a metallized covalent $\sigma$-bond route that is more favorable for achieving high-$T_c$ superconductors, and a Fermi-level density-of-states accumulation route that can enhance $T_c$ but usually to a more limited extent.
\end{abstract}


\maketitle


\section{INTRODUCTION}

The discovery of superconductors with high transition temperatures ($T_c$) remains a major goal in condensed-matter physics and materials design. While unconventional systems such as cuprates, iron-based superconductors, and nickelates have greatly broadened the conceptual landscape of superconductivity~\cite{cup1,cup2,iro1,iro2,iro3,Nick}, phonon-mediated superconductors~\cite{bcs} remain the most accessible class for first-principles predictive investigation. In particular, the recent success of hydrogen-rich compounds~\cite{metalHpre,LaH10_1st,LaH10_2nd,SH3_1st,maprl,maprl2,CaH6_1st,CaH6_2nd,YH9_1st} under extreme pressure has demonstrated that conventional electron-phonon mechanisms can support superconductivity at remarkably high temperatures. These developments have further emphasized the importance of building efficient and reliable computational frameworks for identifying new superconductors, especially those that may be stabilized under less extreme conditions.

For phonon-mediated superconductors, the standard theoretical route is well established. Within density functional perturbation theory (DFPT)~\cite{dfptreview,dfptreview2,pwscf}, one can calculate phonon spectra, electron-phonon coupling (EPC) matrix elements, and the Eliashberg spectral function, and then estimate superconducting $T_c$ using the McMillan-Allen-Dynes (MAD) formula~\cite{mcmillan1,mcmillan2} or the Migdal-Eliashberg self-consistent equation~\cite{epw,eliashberg}. However, despite its accuracy and wide acceptance, this workflow is computationally costly. Converged calculations typically require self-consistent linear-response treatments together with dense sampling of both electronic and phononic Brillouin zones. As a consequence, DFPT-based electron-phonon calculations are far more expensive than ordinary density functional theory (DFT) electronic-structure calculations~\cite{dft1,dft2}, which makes direct large-scale screening across broad materials spaces impractical.

Many studies have therefore explored simpler descriptors that can identify promising superconductors before full EPC calculations are performed. Common indicators such as the Fermi-level density of states, characteristic phonon scales, or empirical elemental rules may offer useful hints, but they do not directly measure the microscopic electron-lattice scattering processes that determine EPC strength~\cite{dfptreview,dfptreview2}. Data-driven and machine-learning approaches provide another possible route~\cite{ML_sc1,ML_sc2}, yet they often face two persistent difficulties: the limited availability of high-quality superconductivity datasets and the lack of clear physical interpretability. Recently, the \(\sigma\)-bonding density of states has been introduced as a descriptor for identifying strong-EPC and potentially high-\(T_c\) materials from machine-learning Hamiltonians~\cite{deepH,sigmados}, highlighting the promise of physically interpretable screening strategies.

In this work, we propose a low-cost framework for estimating the electronic tendency toward strong EPC directly from static orbital-based Hamiltonians, without explicitly carrying out phonon perturbation calculations at the screening stage. The key idea is to infer displacement-induced Hamiltonian variations from static interatomic hybridization information, thereby capturing the leading electronic ingredients of EPC in a computationally efficient manner. Based on this static electron-phonon response (SEPR) scheme, we define two screening quantities, $\Xi$ and $\chi$, which characterize, respectively, the overall scale of EPC-related scattering and the intrinsic deformation response of the electronic structure. Benchmarks against representative superconductors show that these quantities reproduce the correct order of magnitude and main material trends of EPC-related behavior with substantially reduced computational cost.

We further combine this framework with the MattKeyBond database~\cite{MKB}, which contains detailed bond-centric electronic information for more than 36,000 compounds. Based on high-throughput screening using $\Xi$ and $\chi$, followed by rigorous DFPT verification, we identified 34 dynamically stable superconducting candidates with calculated $T_c > 10$ K. Analysis of these candidates reveals two common routes toward relatively high-$T_c$ superconductivity. One route is associated with metallized covalent $\sigma$-bonds, which tend to generate large deformation potentials and are more favorable for reaching high $T_c$. The other route relies on an enhanced density of active electronic states near the Fermi level, which can also strengthen superconductivity, although usually less pronounced. At the same time, the screened materials also show a strong competition between EPC and lattice stability. This competition places an important limit on the search for materials with higher $T_c$.

Overall, our results provide both a practical screening strategy and a physically interpretable picture of how orbital hybridization, bonding character, and band structure cooperate to produce strong electron-phonon coupling and high $T_c$. By avoiding explicit linear-response calculations in the initial search stage, the present approach offers a scalable route for new-type of superconductor discovery across large chemical spaces.

\begin{figure*}[t]
\includegraphics[angle=0,scale=0.6]{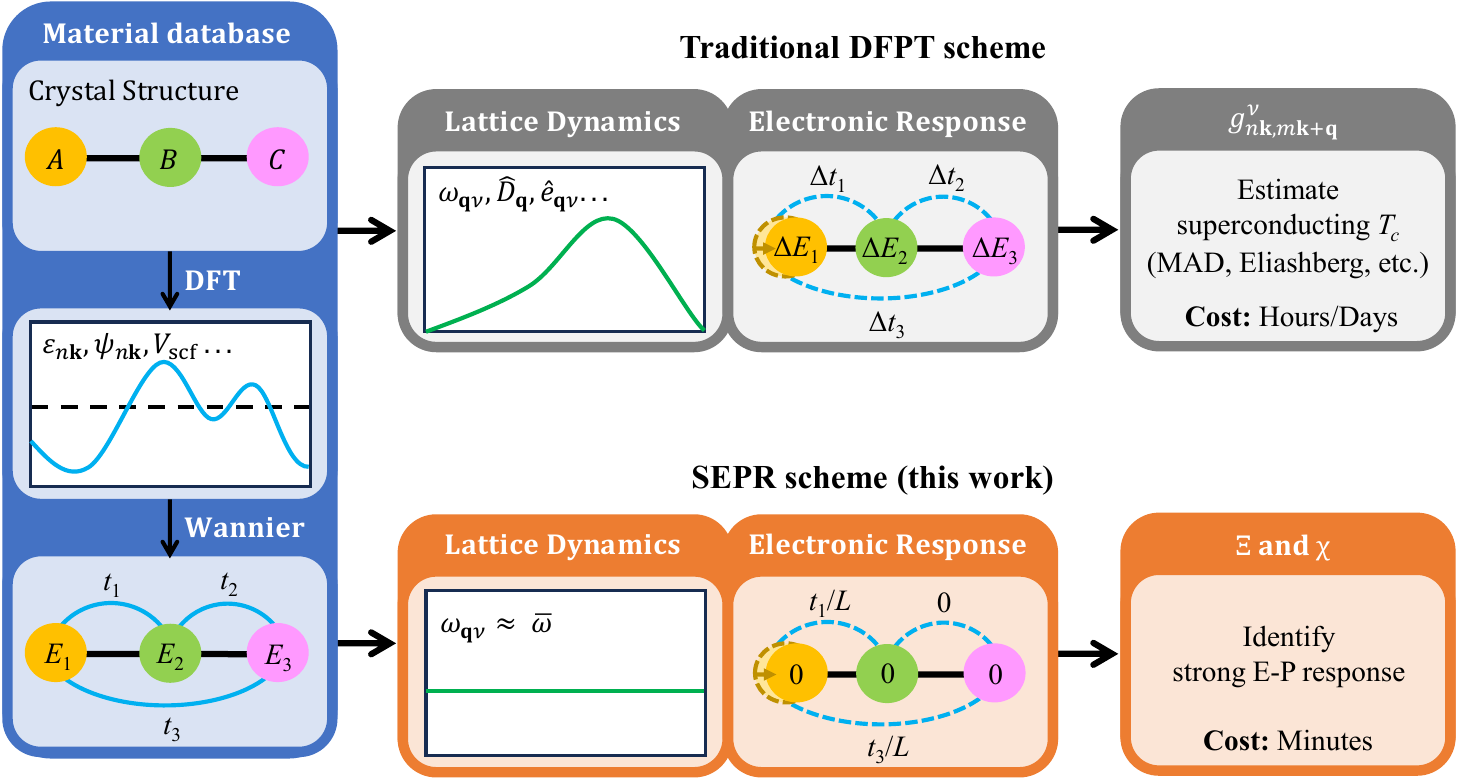}
\caption{
Schematic illustration of the static electron-phonon response (SEPR) scheme framework. The gray panel shows the conventional DFPT-based workflow. The orange panel summarizes the present SEPR workflow, which consists of three successive approximations: (i) the dynamic isotropic approximation, which replaces mode-dependent phonon frequencies by an average characteristic frequency $\bar{\omega}$; (ii) the double-center approximation, which retains the hopping variations associated with bonds connected to the displaced atom; and (iii) the bonding attractivity (BA) model approximation, which estimates the hopping derivatives from the static Hamiltonian through a bond-length dependence. The resulting screening parameters $\Xi$ and $\chi$ characterize the EPC-related electronic response and are used for rapid identification of strong electron-phonon-coupling materials.
}
\label{Fig_outline}
\end{figure*}

\section{EPC estimation from static Hamiltonians}

Our goal is not to reproduce the full DFPT workflow at the screening stage, but to construct low-cost quantities that retain the leading electronic trends governing strong EPC. In phonon-mediated superconductors, the total EPC strength is determined jointly by the electron/phonon spectra and the electron-phonon scattering matrix. Among these ingredients, the most difficult part for large-scale screening is the explicit evaluation of phonon-displacement induced perturbation Hamiltonians and their integration over dense electron and phonon momenta. We therefore seek a simpler description that extracts the electronic tendency for strong EPC directly from a static orbital-based Hamiltonian.

Within the Migdal approximation~\cite{epw,eliashberg}, the total EPC constant can be written as
\begin{equation}
\label{eq_lam}
\lambda = \sum_{\mathbf{q}\nu} \lambda_{\mathbf{q}\nu} = \sum_{\mathbf{q}\nu} \frac{\Pi''_{\mathbf{q}\nu}}{\pi N(\varepsilon_F)\hbar^2\omega_{\mathbf{q}\nu}^2},
\end{equation}
where \(N(\varepsilon_F)=\sum_{n\mathbf{k}}\delta(\varepsilon_{n\mathbf{k}}-\varepsilon_F)\) is the electronic density of states (DOS) at the Fermi level, \(\omega_{\mathbf{q}\nu}\) is the phonon frequency of mode \((\mathbf{q},\nu)\), and \(\Pi''_{\mathbf{q}\nu}\) is the imaginary part of the phonon self-energy within electron-phonon interaction. Within the double-\(\delta\) approximation~\cite{epw,eliashberg},
\begin{eqnarray}
\Pi''_{\mathbf{q}\nu}  &=& \pi\hbar^2 \sum_{\mathbf{k}nm} \delta(\varepsilon_{n\mathbf{k}}-\varepsilon_F) \delta(\varepsilon_{m\mathbf{k+q}}-\varepsilon_F) \nonumber\\
&&\sum_{A\alpha,B\beta} \frac{e^{A\alpha*}_{\mathbf{q}\nu} e^{B\beta}_{\mathbf{q}\nu}}{2\sqrt{m_A m_B}} \nonumber \\
&&\langle n \mathbf{k}| \frac{d\widehat{H}^\dagger}{du_{A\alpha}(\mathbf{q})} |m \mathbf{k} + \mathbf{q}\rangle \nonumber \\
&& \langle m \mathbf{k} + \mathbf{q}| \frac{d\widehat{H}}{du_{B\beta}(\mathbf{q})} |n \mathbf{k}\rangle.
\label{eq_Pi}
\end{eqnarray}

Equation \ref{eq_Pi} makes clear that the central electronic quantity controlling EPC is the displacement derivative of the Hamiltonian, \(d\widehat H/du\). In DFPT, this object is obtained from self-consistent linear-response calculations of the density, potential, and Kohn-Sham wavefunctions~\cite{dfptreview,dfptreview2}. That procedure is accurate, but it is also the main computational bottleneck. The strategy of the present work is therefore to approximate the dominant electronic part of \(d\widehat H/du\) from static interatomic hybridization information.

To this end, we introduce a static electron-phonon response (SEPR) scheme, schematically shown in Fig. \ref{Fig_outline}. The construction consists of three approximations. First, we average over the detailed phonon-mode dependence and extract an overall EPC-related scattering scale. Second, we approximate the displacement-induced Hamiltonian variation by retaining only the bonds directly connected to the displaced atom. Third, we estimate the resulting hopping derivatives from the bond-length dependence of the static orbital Hamiltonian. These three steps lead to two screening quantities, \(\Xi\) and \(\chi\), which characterize the mass-weighted EPC-related scattering scale and the intrinsic electronic deformation tendency, respectively.

\subsection{Dynamic Isotropic Approximation}

We first remove the detailed mode dependence of the phonon spectrum by replacing the EPC-active phonon frequencies with a characteristic average value, \(\omega_{q\nu}\approx \bar\omega\). Under this dynamic isotropic approximation, the total EPC constant can be rewritten as
\begin{equation}
\lambda \approx \frac{\Xi^2}{\hbar^2\bar{\omega}^2}, \qquad
\Xi^2 = \sum_{\mathbf{q}\nu} \frac{\Pi_{\mathbf{q}\nu}^{\prime\prime}} {\pi N(\varepsilon_F)}.
\label{eq_lam2}
\end{equation}
The quantity \(\Xi\) has the dimension of energy and measures the overall scale of EPC-related electronic scattering after the detailed phonon-frequency distribution has been factored out. It is introduced here as a screening parameter, not as a replacement for a full calculation of \(\lambda\), \(\alpha^2F(\omega)\), or \(T_c\). 

\begin{figure}[t]
\includegraphics[angle=0,scale=0.48]{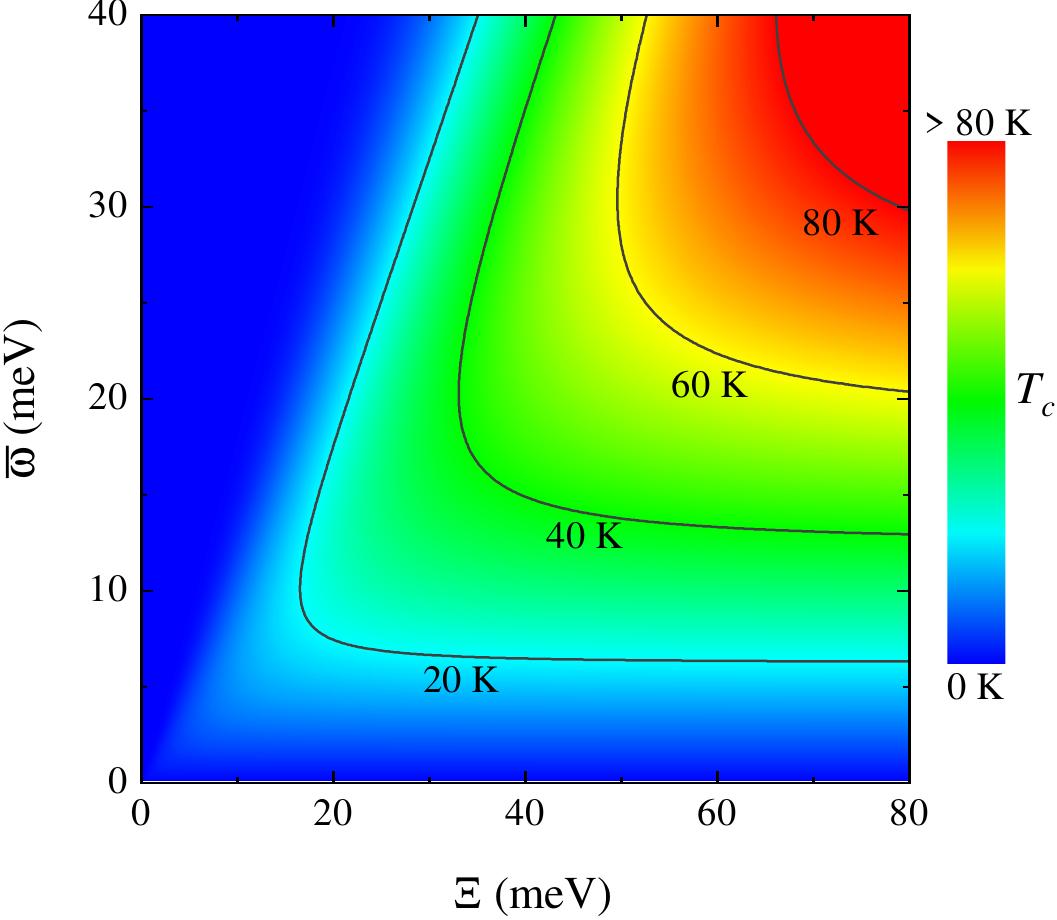}
\caption{
Contour map of the superconducting transition temperature \(T_c\) in the \(\Xi-\bar{\omega}\) plane, evaluated from the McMillan-Allen-Dynes formula. Within dynamic isotropic approximation, we have $\omega_\text{log}=\bar\omega$ and \(\lambda=\Xi^2/\hbar^2\bar{\omega}^2\). \(\mu^{\ast}\) was empirically set as 0.1. 
}
\label{Fig_MAD}
\end{figure}

This distinction is important because phonon energy enters superconductivity in two competing ways. Within equation \ref{eq_lam2}, increasing \(\bar\omega\) suppresses \(\lambda\) through the factor \(1/\bar\omega^2\). However, if the lattice is too soft, the characteristic energy scale for Cooper pairing also becomes small. This competition effect has been reported in several pressure-insensitive superconducting systems\cite{Nb_alloy1,Nb_alloy2,NbTi,MoB2,RSAVS}. The corresponding balance is illustrated by the contour map of McMillan-Allen-Dynes formula~\cite{mcmillan1,mcmillan2} in Fig. \ref{Fig_MAD}. Where a large \(\Xi\) cannot by itself guarantee a high \(T_c\) if \(\bar\omega\) is too high or too low, but a sufficiently large \(\Xi\) is still a necessary indicator of strong EPC. For screening purposes, \(\Xi\) therefore serves as an initial measure of whether a material has the electronic capacity to support high superconducting \(T_c\).

\subsection{Double-center Approximation}

To evaluate \(\Xi\) without explicit DFPT calculations, we represent the electronic structure by the orbital-based tight-binding Hamiltonian \(H_{Aa,Bb}\) constructed in the MattKeyBond database~\cite{MKB}, where \(a\) and \(b\) denote Closest Wannier Function (CWF) orbitals~\cite{cwf} centered on atoms \(A\) and \(B\). The remaining challenge is then to estimate the atom-displacement derivative \(d\widehat H/du\).

To make this quantity tractable within a static-Hamiltonian framework, we adopt the double-center approximation~\cite{NP_topo}.
As illustrated in Fig. \ref{Fig_outline}, when an atom \(A\) is displaced, the dominant variation of \(\widehat H\) is assumed to arise from the hopping terms between that atom and its directly bonded neighbors, while the variations in onsite terms and higher-order multicenter terms are neglected. 

Combining the double-center approximation with the dynamic isotropic approximation, one can recast the expression for \(\Xi\) into a form involving only electronic quantities:
\begin{eqnarray}
\Xi^2 &&= \sum_{A\alpha} \Xi^2_{A\alpha} \nonumber\\
&&= \frac{\hbar^2}{N(\varepsilon_F)} \text{Re} \sum_{A\alpha} \frac{1}{m_A} \text{tr}\Big(\widehat{G}_{A\alpha}\widehat{G}_{A\alpha} + \widehat{M}\widehat{F}_{A\alpha,A\alpha}\Big),
\label{eq_Xi}
\end{eqnarray}
where \(\widehat M\), \(\widehat G_{A\alpha}\), and \(\widehat F_{A\alpha,A\alpha}\) are auxiliary matrices obtained from a single \(\mathbf{k}\)-space summation, and their explicit forms are given in the Supplemental Material. Here \(\widehat M\) describes the electronic density matrix near the Fermi level, whereas \(\widehat G_{A\alpha}\) and \(\widehat F_{A\alpha,A\alpha}\) contain the information associated with the hopping derivatives \(d\widehat H/du_{A\alpha}\).

Equation \ref{eq_Xi} shows that \(\Xi\) can be decomposed naturally into atom- and direction-resolved contributions. This decomposition is useful not only for screening, but also for microscopic interpretation, because it directly identifies which atoms and which displacement directions dominate the EPC-related response.

Although \(\Xi\) is closely connected to the overall EPC scale, it still contains the ionic mass factor \(1/m_A\). As a result, a screening strategy based on \(\Xi\) alone may favor light-element systems even when their phonon frequencies fall outside the optimal range for superconductivity. To isolate the purely electronic deformation tendency, we introduce a second screening quantity,
\begin{eqnarray}
\chi^2 &&= \sum_{A\alpha} \chi^2_{A\alpha} \nonumber\\
&&= \frac{\hbar^2}{N(\varepsilon_F)}\text{Re} \sum_{A\alpha} \text{tr}\Big(\widehat{G}_{A\alpha}\widehat{G}_{A\alpha} + \widehat{M}\widehat{F}_{A\alpha,A\alpha}\Big) .
\label{eq_chi}
\end{eqnarray}
Relative to \(\Xi\), the quantity \(\chi\) removes the mass weighting and therefore depends only on the electronic structure and orbital hybridization. The two quantities thus play complementary roles: \(\Xi\) is more directly connected to the total EPC scale, whereas \(\chi\) characterizes the intrinsic electronic deformation response independent of ionic mass.

\subsection{Bonding Attractivity Model Approximation}

The final step is to estimate the hopping derivatives entering Eqs. \ref{eq_Xi} and \ref{eq_chi} directly from the static Hamiltonian. This is the key approximation that allows the SEPR framework to bypass self-consistent linear-response calculations.

We assume that the hopping integral between two orbitals decays approximately exponentially with bond length,
\begin{equation}
H_{ab}(R) \propto \text{exp}(-R/L),
\end{equation}
where \(L\) is a characteristic decay length. This assumption is motivated by the bond-length dependence of the interatomic ICOHP revealed by the Bonding Attractivity (BA) model in the MattKeyBond framework~\cite{MKB}. Physically, the ICOHP between atoms $A$ and $B$ can be defined as the trace of the product of the Hamiltonian and the density matrix \(\widehat D\), i.e., \(\text{ICOHP}_{AB} = \sum_{a \in A, b \in B} H_{ab}D_{ba}\). Within the Hellmann–Feynman framework, the first-order change in the electronic density does not contribute explicitly to the first-order variation of the total energy. At the present level of approximation, we therefore assign the leading bond-length dependence of the bond energy to the hopping matrix.

Under this assumption, the displacement derivative of the hopping matrix can be approximated as
\begin{equation}
\frac{dH_{ab}}{du} \approx -\frac{1}{L}\frac{\mathbf{R}\cdot\mathbf{e}}{|\mathbf{R}|}H_{ab}(\mathbf{R}),
\label{eq_dhdu}
\end{equation}
where \(\mathbf R\) is the bond vector and \(\mathbf e\) is the displacement direction. The magnitude of this derivative is maximized when the displacement is parallel or antiparallel to the bond direction, giving \(\pm H_{ab}/L\). Equation \ref{eq_dhdu} therefore provides a direct real-space interpretation of strong EPC tendency: strongly hybridized bonds around the Fermi level, especially directional \(\sigma\) bonds, naturally generate larger deformation responses and higher superconducting $T_c$.

In practical high-throughput screening, we use a uniform BA decay length \(L_A=1.2\) Å for all elements. Under Eq. \ref{eq_dhdu}, the hopping derivative scales approximately as \(dH_{ab}/du \sim H_{ab}/L\), and therefore both \(\Xi\) and \(\chi\) inherit an approximate inverse dependence on the effective decay length. This choice provides a simple and consistent numerical yardstick across large chemical space, while also introducing a predictable family-dependent bias when the true decay length differs substantially from the reference value. We analyze this systematic bias explicitly in Sec. V.

Taken together, the three approximations above convert the EPC problem from an explicit lattice-dynamical calculation into a set of low-cost screening parameters that can be evaluated directly from a static orbital Hamiltonian. The resulting quantities, \(\Xi\) and \(\chi\), do not replace DFPT for quantitative predictions on individual compounds, but they retain the dominant electronic trends relevant for strong EPC and therefore make large-scale screening feasible.

\section{VERIFICATION USING REPRESENTATIVE SUPERCONDUCTORS}

\begin{table*}[tb]
\caption{
Comparison of the screening parameters \(\Xi\) and \(\chi\) obtained from the SEPR scheme and from DFPT for representative phonon-mediated superconductors. For each material, the table lists the total values of \(\Xi\) and \(\chi\), together with atom-resolved contributions. The corresponding computational time costs under the same computing setup are also included. Here \(\Xi\) is given in meV and \(\chi\) in meV\(\cdot\)u\(^{1/2}\).
}
\begin{center}
\begin{tabular*}{16cm}{@{\extracolsep{\fill}} cccccccc}

\hline\hline															
System	&	Component	&	$\Xi$ (meV)	&	$\Xi$ (meV)	&	$\chi$ (meV$\cdot$u$^{1/2}$)	&	$\chi$ (meV$\cdot$u$^{1/2}$)	&	Time cost	&	Time cost	 \\
	&		&	SEPR	&	DFPT	&	SEPR	&	DFPT	&	SEPR	&	DFPT	 \\
\hline															
MgB$_2$	&	Total	&	43.3 	&	55.9 	&	152.5 	&	186.7 	&	1.5 m	&	2.2 h	 \\
	&	Mg	&	14.8 	&	8.8 	&	72.8 	&	43.3 	&		&		 \\
	&	B	&	40.8 	&	55.2 	&	134.0 	&	181.6 	&		&		 \\
\hline															
BaBiO$_3$	&	Total	&	33.2 	&	39.0 	&	141.8 	&	172.1 	&	0.7 m	&	0.9 h	 \\
	&	Ba	&	0.5 	&	2.2 	&	5.2 	&	25.2 	&		&		 \\
	&	Bi	&	3.6 	&	5.0 	&	51.4 	&	71.6 	&		&		 \\
	&	O	&	33.0 	&	38.6 	&	132.1 	&	154.4 	&		&		 \\
\hline															
Nb$_3$Ge	&	Total	&	22.7 	&	24.9 	&	216.8 	&	238.7 	&	20.3 m	&	5.0 h	 \\
	&	Nb	&	21.6 	&	24.1 	&	208.1 	&	232.7 	&		&		 \\
	&	Ge	&	7.1 	&	6.3 	&	60.7 	&	53.2 	&		&		 \\
\hline															
Nb	&	Total	&	22.8 	&	19.1 	&	219.4 	&	183.6 	&	2.7 m	&	7.5 h	 \\
\hline															
Al	&	Total	&	22.9 	&	19.3 	&	118.8 	&	100.2 	&	0.5 m	&	7.6 h	 \\
\hline															
SH$_3$	&	Total	&	181.6 	&	195.4 	&	252.4 	&	278.6 	&	2.7 m	&	8.0 h	 \\
	&	S	&	31.3 	&	35.5 	&	177.4 	&	201.0 	&		&		 \\
	&	H	&	178.9 	&	192.1 	&	179.6 	&	192.9 	&		&		 \\
\hline															
LaH$_{10}$	&	Total	&	177.6 	&	213.1 	&	192.7 	&	249.3 	&	61.3 m	&	26.5 h	 \\
	&	La	&	6.2 	&	10.9 	&	73.3 	&	128.5 	&		&		 \\
	&	H	&	177.5 	&	212.8 	&	178.1 	&	213.6 	&		&		 \\
\hline\hline															

\end{tabular*}
\end{center}
\label{tab_bench}
\end{table*}

Before applying SEPR to large-scale screening, we first test whether it provides a reliable low-cost proxy for DFPT in known phonon-mediated superconductors. Our benchmark set, summarized in Table \ref{tab_bench}, spans representative superconductors with diverse bonding and structural characters, including MgB$_2$~\cite{MgB2_1st}, BaBiO$_3$ (perovskite phase) as the parent compound of superconducting Ba$_{1-x}$K$_x$BiO$_3$~\cite{bkbo_pre,bkbo}, the A15 compound Nb$_3$Ge~\cite{Nb3Ge}, elemental Nb and Al~\cite{sc_sum}, and the high-pressure hydrides SH$_3$ and LaH$_{10}$~\cite{LaH10_1st,LaH10_2nd}. As shown in Table \ref{tab_bench}, despite the multiple approximations introduced in Sec. II, SEPR method reproduces the correct order of magnitude of \(\Xi\) and \(\chi\) and captures the main trends across diverse material families. This level of agreement is sufficient for screening, where the central question is whether a material has a strong intrinsic tendency toward EPC and thus deserves subsequent DFPT verification.

A further application of SEPR is that both \(\Xi\) and \(\chi\) allow atom- and direction-resolved decompositions in Eqs. \ref{eq_Xi} and \ref{eq_chi}. This enables a microscopic validation of the method beyond total scalar comparisons (see the Supplemental Material for the detailed decomposition). In MgB$_2$, for example, the dominant contributions come from in-plane B displacements rather than Mg, fully consistent with the established picture of metallized B-B \(\sigma\) bonds strongly coupled to the bond-stretching modes~\cite{MgB2_cal,MgB2_sig,MgB2_arpes}. Likewise, for the bismuthate system, although a fully quantitative treatment of superconductivity remains delicate within semilocal PBE-based calculations~\cite{PBE}, SEPR correctly highlights the Bi-O network and oxygen motions along the Bi-O bonds as the dominant channel, in line with the accepted physical mechanism~\cite{bkbo_cal1,bkbo_cal2}. By contrast, in A15 compounds and elemental superconductors, the contributions are more broadly distributed, reflecting their more metallic and less strongly anisotropic bonding and EPC characters~\cite{A15,Nb3Sn_cal,Nb_cal}. 

The benchmark systems also clarify the complementary roles of \(\Xi\) and \(\chi\). Because \(\Xi\) retains the ionic-mass weighting, it is more directly connected to the overall EPC scale, whereas \(\chi\) isolates the intrinsic electronic deformation response. Elemental Al illustrates this distinction clearly: although its \(\Xi\) is comparable to that of Nb and several A15 compounds, its \(T_c\) is much lower~\cite{A15,Nb3Sn_cal,Nb_cal}. This contrast arises because the low atomic mass of Al enhances $\Xi$, while its smaller \(\chi\) reflects a weaker intrinsic electronic response to lattice. The hydrides SH$_3$ and LaH$_{10}$ provide the opposite example. Although their very large \(\Xi\) values are partly associated with the light mass of hydrogen, their large \(\chi\) also indicate intrinsically strong electron-lattice responses. This is consistent with our previous finding that hydrogen owns the strongest bonding ability via covalent hybridization~\cite{MKB}.

Finally, Table \ref{tab_bench} shows the key practical advantage of SEPR: computational efficiency. Standard DFPT calculations require explicit phonon perturbations together with dense integrations over both electron and phonon momenta~\cite{dfptreview,dfptreview2}, which makes EPC calculations far more expensive than ordinary DFT electronic-structure calculations. Even for relatively simple systems such as MgB$_2$ and BaBiO$_3$, the DFPT calculations already require hours under our computing setup, and the cost rises rapidly for larger or lower-symmetry structures. By contrast, the corresponding SEPR evaluations are completed within minutes under the same computational resources (96 CPUs per task, see the Supplemental Material for computational details). This dramatic reduction in computational cost removes the main bottleneck that has long limited large-scale EPC screening, shifting EPC-oriented materials discovery from expensive one-by-one calculation strategy to a realistic high-throughput workflow.

Overall, the benchmark results show that SEPR captures the correct order of magnitude of $\Xi$ and $\chi$ at a fraction of the cost of DFPT, while preserving a clear microscopic interpretation. This justifies the feasibility of the large-scale screening presented below.

\section{HIGH-THROUGHPUT SCREENING}

\begin{figure}[t]
\includegraphics[angle=0,scale=0.52]{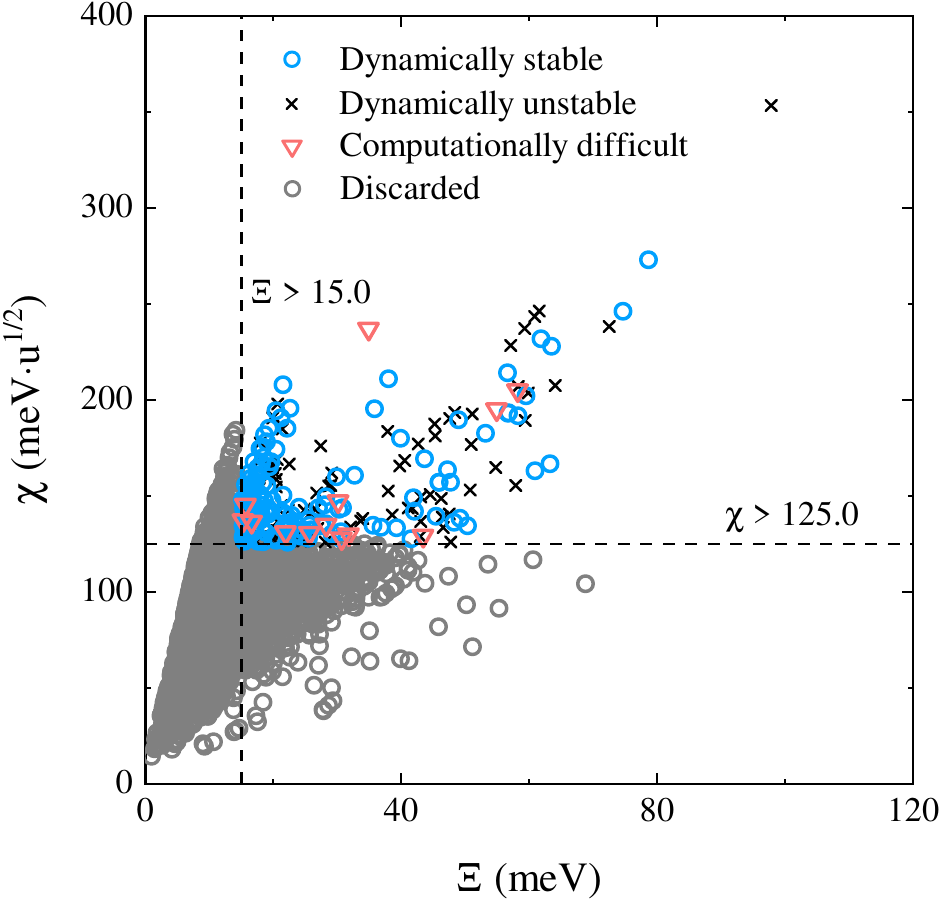}
\caption{
The calculated \(\Xi\) and \(\chi\) values for the 8,833 nonmagnetic metallic compounds. The dashed lines indicate the empirical screening thresholds, \(\Xi = 15.0\) meV and \(\chi = 125.0\) meV\(\cdot\)u\(^{1/2}\). Different symbols distinguish dynamically stable compounds, dynamically unstable compounds, computationally difficult cases, and discarded materials.
}
\label{Fig_workflow}
\end{figure}

We now apply the SEPR scheme to the large-scale search for phonon-mediated superconductors in the MattKeyBond database~\cite{MKB}. The starting dataset contains the bond-centric electronic information for 36,377 compounds collected from the Materials Project (MP)~\cite{mp} and the Inorganic Crystal Structure Database (ICSD)~\cite{icsd}. Because the present framework is designed for phonon-mediated superconductivity, we first retain only metallic compounds and then exclude magnetic systems, leaving 8,833 nonmagnetic metals for SEPR screening. For each of these materials, we evaluate the screening quantities \(\Xi\) and \(\chi\). As shown in Fig.~\ref{Fig_workflow}, most compounds are concentrated in the low-\(\Xi\), low-\(\chi\) region, illustrating the rarity of strong EPC propensity in broad chemical space. Using the benchmarked criteria \(\Xi > 15.0\) meV and \(\chi > 125.0\) meV\(\cdot \text{u}^{1/2}\), we reduce the candidate pool to 260 compounds. 

These preselected materials are then examined by full structural relaxation and DFPT phonon/EPC calculations. Of the 260 compounds, 150 are dynamically stable. Their superconducting transition temperatures are subsequently estimated using the McMillan-Allen-Dynes formula~\cite{mcmillan1,mcmillan2}, yielding 99 systems with \(T_c > 2\) K and 34 with \(T_c > 10\) K. The substantial loss of candidates during dynamical-stability verification points to a recurring competition between strong EPC and lattice stability, which will be analyzed further in Sec. VI.

The final list of 34 systems with \(T_c > 10\) K, summarized in Table~\ref{tab_list}, contains both well-known phonon-mediated superconductors and previously overlooked candidates. The former include representative benchmark materials such as MgB\(_2\), BaBiO\(_3\), and several A15 compounds, which provides an internal consistency check for the entire workflow. More importantly, the screened set also contains several materials that have already been synthesized but, to our knowledge, still lack low-temperature transport characterization, as well as several structure-predicted compounds without experimental synthesis records. These materials therefore constitute promising targets for future experimental and theoretical follow-up. Detailed structural, electronic, phononic, and Eliashberg spectral data for the verified candidates are provided in the Supplemental Material.

Taken together, these results show that screening based on \(\Xi\) and \(\chi\) can effectively identify strong-EPC superconductors without explicit phonon calculations at the initial stage. The SEPR scheme therefore serves as an efficient and physically interpretable filter for further superconductor discovery.

\begin{table*}[!p]
\caption{
List of the 34 screened dynamically stable superconducting candidates with experimental or calculated \(T_c > 10\) K. The candidates are grouped according to their current status: experimentally examined materials and our newly identified superconducting candidates with and without experimental synthetic records. For each compound, the table lists the Materials Project ID, chemical formula, the screening parameters \(\Xi\) and \(\chi\) obtained from the SEPR and DFPT calculations, and the superconducting transition temperature \(T_c\). Here \(\Xi\) is in meV and \(\chi\) is in meV\(\cdot\)u\(^{1/2}\). Some materials are only slightly dynamically unstable at ambient pressure but become stable under applied pressure. Superscripts on their \(T_c\) denote the pressure conditions. 
}
\label{tab_list}
\centering

\vspace{1.0em}
\noindent\makebox[16cm][l]{\text{{\bf{Experimentally examined}}}}
\begin{tabular*}{16cm}{@{\extracolsep{\fill}}lcccccc@{}}												
\hline \hline															
mp. ID.	&	Formula	&	$\Xi$ (SEPR)	&	$\Xi$ (DFPT)	&	$\chi$ (SEPR)	&	$\chi$ (DFPT)	&	$T_c$ (exp.)	\\
\hline													
mp-763	&	MgB$_2$	&	43.3 	&	55.9 	&	152.5	&	186.7 	&	39.0 	\\
mp-545783	&	BaBiO$_3$	&	33.2 	&	39.0 	&	141.8	&	172.1 	&	31.0 	\\
mp-1373	&	Nb$_3$Ge	&	22.7 	&	24.9 	&	216.8	&	238.7 	&	23.2 	\\
mp-10229	&	Nb$_3$Si	&	21.2 	&	23.9 	&	190.6 	&	216.2 	&	19.0 	\\
mp-1326	&	Nb$_3$Sn	&	18.0 	&	23.6 	&	174.5 	&	228.3 	&	18.1 	\\
mp-796	&	Nb$_3$Al	&	22.6 	&	25.3 	&	195.5 	&	233.8 	&	18.0 	\\
mp-2670	&	Nb$_3$Ga	&	20.5 	&	24.6 	&	194.3 	&	236.0 	&	14.5 	\\
mp-571664	&	Zr$_2$Rh	&	15.4 	&	15.9 	&	148.1 	&	154.8 	&	11.2 	\\
mp-29239	&	LuB$_2$Ru	&	30.9 	&	36.0 	&	143.7 	&	182.5 	&	10.0 	\\

\hline \hline																	
\end{tabular*}

\vspace{1.0em}
\noindent\makebox[16cm][l]{\text{{\bf{Newly identified, ``with" experimental synthetic record}}}} 
\begin{tabular*}{16cm}{@{\extracolsep{\fill}}lcccccc@{}}												
\hline \hline													
mp. ID.	&	Formula	&	$\Xi$ (SEPR)	&	$\Xi$ (DFPT)	&	$\chi$ (SEPR)	&	$\chi$ (DFPT)	&	$T_c$ (cal.)	\\
\hline													
mp-576	&	B$_{13}$C$_2$	&	74.7 	&	71.7 	&	246.1 	&	236.6 	&	41.2 	\\
mp-1102261	&	IrS$_2$	&	22.5 	&	23.9 	&	135.8 	&	178.7 	&	17.3 	\\
mp-15660	&	Nb$_4$C$_3$	&	18.7 	&	35.9 	&	147.4 	&	207.3 	&	16.4$^{10.0}$	\\
mp-1432	&	Be$_2$B	&	45.5 	&	48.4 	&	139.3 	&	149.4 	&	16.3 	\\
mp-11271	&	BeMo$_3$	&	22.9 	&	23.2 	&	139.8 	&	192.4 	&	16.0 	\\
mp-12760	&	Zr$_6$Al$_2$Co	&	19.7 	&	15.1 	&	167.7 	&	130.4 	&	15.1 	\\
mp-1103173	&	Nb$_4$FeP	&	16.8 	&	19.0 	&	150.6 	&	161.1 	&	14.0 	\\
mp-1189682	&	LuB$_2$C	&	27.6 	&	45.5 	&	134.4 	&	177.8 	&	13.5 	\\
mp-92	&	Si	&	27.1 	&	24.1 	&	143.7 	&	127.8 	&	12.7 	\\
mp-11038	&	Zr$_6$Al$_2$Fe	&	19.5 	&	15.6 	&	165.1 	&	133.4 	&	11.8 	\\
mp-30619	&	Zr$_3$Co	&	17.6 	&	15.4 	&	164.4 	&	139.8 	&	10.9 	\\
mp-1061054	&	Ge	&	15.9 	&	14.4 	&	135.9 	&	122.4 	&	10.6 	\\
mp-1095672	&	Ta$_2$V$_3$Si	&	18.1 	&	20.4 	&	137.4 	&	155.0 	&	10.4 	\\
mp-9459	&	Y$_4$C$_5$	&	24.4 	&	33.4 	&	128.9 	&	142.2 	&	10.4 	\\
mp-1103	&	LuSi$_2$	&	24.1 	&	23.6 	&	144.0 	&	145.6 	&	10.0$^{5.0}$	\\

\hline\hline													
\end{tabular*}

\vspace{1.0em}
\noindent\makebox[16cm][l]{\text{{\bf{Newly identified, ``without" experimental synthetic record}}}} 
\begin{tabular*}{16cm}{@{\extracolsep{\fill}}lcccccc@{}}												
\hline \hline													
mp. ID.	&	Formula	&	$\Xi$ (SEPR)	&	$\Xi$ (DFPT)	&	$\chi$ (SEPR)	&	$\chi$ (DFPT)	&	$T_c$ (cal.)	\\
\hline													
mp-1080030	&	BC$_7$	&	56.7 	&	89.6 	&	192.9 	&	306.8 	&	57.4 	\\
mp-1103047	&	LiBeB	&	49.2 	&	50.8 	&	138.4 	&	160.1 	&	46.0 	\\
mp-1079661	&	BC$_7$	&	56.8 	&	86.8 	&	193.2 	&	297.3 	&	39.0 	\\
mp-1018649	&	BC$_5$	&	59.5 	&	77.1 	&	202.2 	&	262.8 	&	35.4 	\\
mp-1077743	&	BC$_5$	&	46.0 	&	78.9 	&	157.0 	&	270.8 	&	32.3 	\\
mp-1077125	&	BC$_5$	&	39.3 	&	68.7 	&	133.4 	&	232.9 	&	28.4 	\\
mp-1079046	&	BC$_7$	&	46.0 	&	78.5 	&	157.0 	&	269.5 	&	24.3 	\\
mp-999439	&	Nb$_3$Cr	&	18.2 	&	17.2 	&	166.7 	&	152.2 	&	22.1 	\\
mp-12892	&	CaSi$_2$	&	29.9 	&	25.9 	&	159.8 	&	139.4 	&	14.1$^{5.0}$	\\
mp-8635	&	Zr	&	18.4 	&	15.0 	&	175.6 	&	142.8 	&	11.0 	\\

\hline \hline		
\end{tabular*}

\end{table*}

\section{SYSTEMATIC BIAS AND ITS PHYSICAL ORIGIN}

\begin{figure}[t]
\includegraphics[angle=0,scale=0.52]{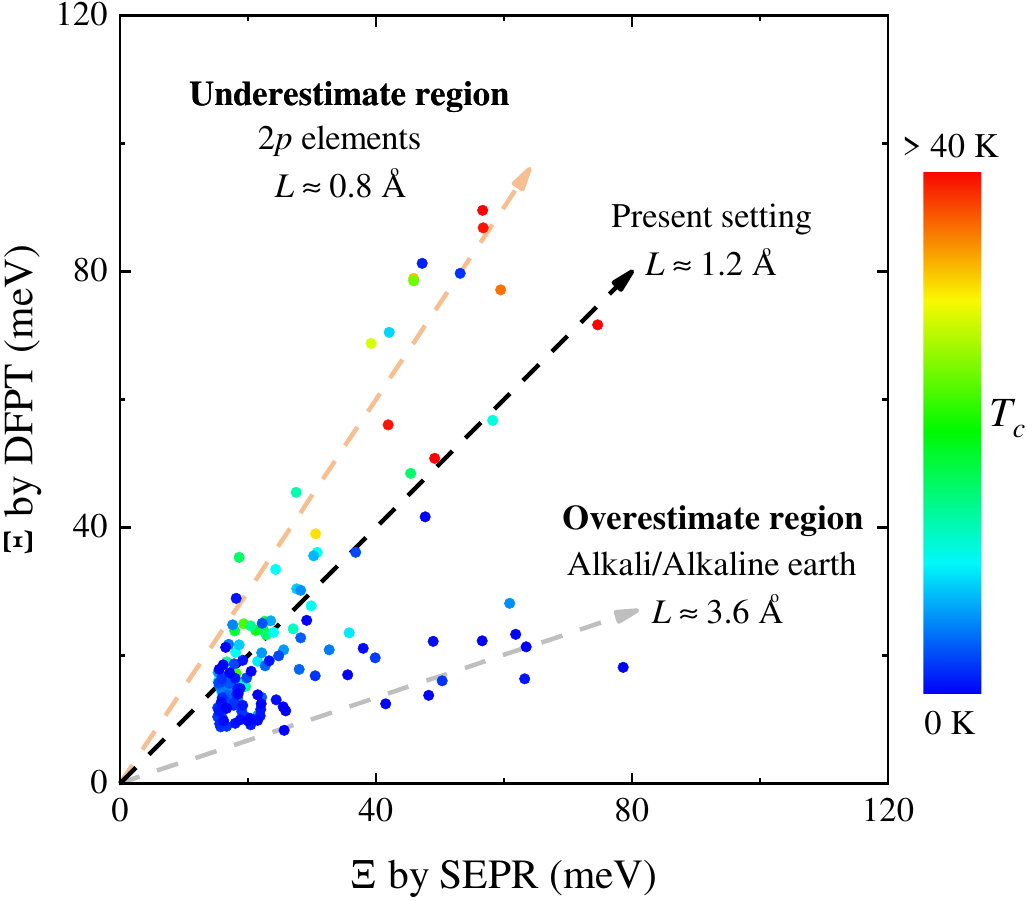}
\caption{
Comparison of the screening parameter \(\Xi\) obtained from the SEPR scheme and from DFPT for the 150 dynamically stable screened compounds. The horizontal axis shows \(\Xi\) from SEPR, and the vertical axis shows \(\Xi\) from DFPT. The dashed diagonal line corresponds to \(\Xi_{\mathrm{SEPR}}=\Xi_{\mathrm{DFPT}}\). Symbol colors indicate the superconducting \(T_c\) (exp. or cal.). 
}
\label{Fig_XivsXi}
\end{figure}

To assess the numerical reliability of SEPR, we compare the screening parameter \(\Xi\) obtained from SEPR with its DFPT counterpart for the 150 dynamically stable screened compounds. As shown in Fig. \ref{Fig_XivsXi}, the two are clearly positively correlated, indicating that SEPR captures the main trend of EPC-related electronic response across a broad range of materials. The deviations, however, are not random: they show a pronounced family dependence, pointing to a systematic bias in the present implementation (see also the full material list in Supplemental material).

The dominant source of this bias originates from the BA-model approximation in Sec. II C. From Eq. \ref{eq_dhdu}, \(dH_{ab}/du \sim H_{ab}/L\), so both \(\Xi\) and \(\chi\) inherit an approximate inverse dependence on the effective decay length \(L\). In the present high-throughput implementation, a uniform value \(L=1.2\) \AA{} is used for all elements and bonding environments. This simplification provides a consistent screening metric across large chemical space, but it inevitably overestimates or underestimates the electron-phonon response for material families whose actual decay lengths differ substantially from the reference value.

This trend is visible in Fig. \ref{Fig_XivsXi}. Compounds containing alkali or alkaline-earth elements are found mainly in the overestimated region, whereas several systems dominated by directional \(2p\)-covalent bonding fall into the underestimated region. These trends are consistent with, respectively, larger and smaller effective decay lengths than the present reference setting. Figure \ref{Fig_XivsXi} should therefore be interpreted not only as a verification plot, but also as a direct visualization of the leading elemental bias in the current SEPR framework.

Additional scatter arises from numerical limitations in both methods. On the SEPR side, the CWF-based Hamiltonian interpolation may retain insufficiently converged or truncated long-range hoppings under finite supercells and finite \(\mathbf{k}\)-mesh resolution. On the DFPT side, the reference EPC quantities can also remain sensitive to the \(\mathbf{k}/\mathbf{q}\)-mesh density and electronic smearing, especially for compounds with sharp Fermi-surface features.

Overall, the comparison in Fig. \ref{Fig_XivsXi} shows that SEPR is sufficiently reliable to capture the main material trends of EPC strength. Its leading deviation is systematic, physically interpretable, and in principle improvable, which makes it well suited for large-scale pre-screening of strong-EPC superconductors.

\section{DUAL PATHWAYS TO HIGH-\(T_c\) EPC SUPERCONDUCTIVITY}

\begin{figure*}[t]
\includegraphics[angle=0,scale=0.55]{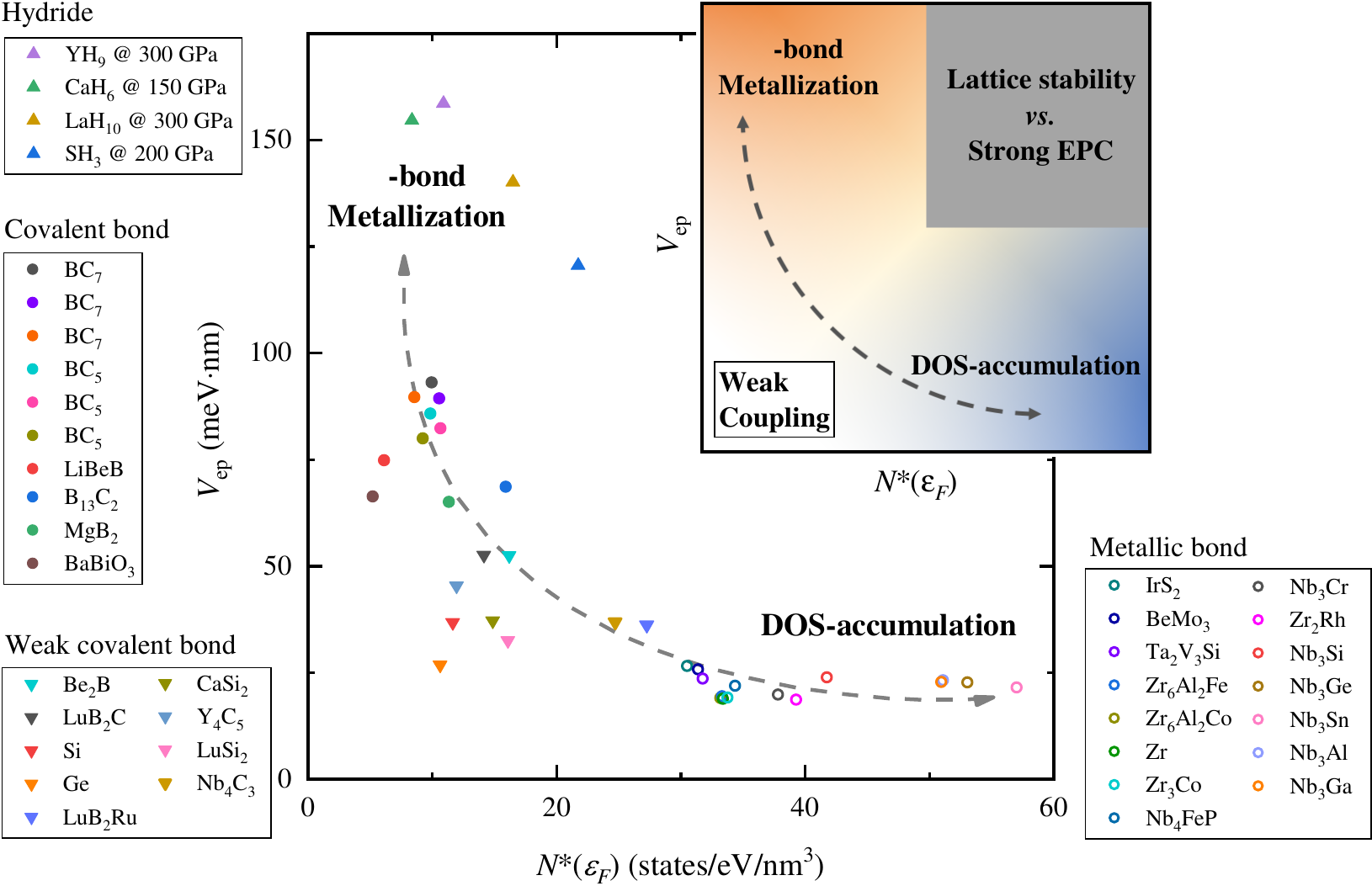}
\caption{
Distribution of the 34 screened dynamically stable superconducting candidates with calculated \(T_c > 10\) K, together with several representative high-pressure hydrides, in the \(V_\text{ep}-N^{\ast}(\varepsilon_F)\) plane. The horizontal axis is the active density of states per unit volume, \(N^{\ast}(\varepsilon_F)\), and the vertical axis is the average electron-phonon scattering measure, \(V_\text{ep}\). Different symbols denote different material classes, including high-pressure hydrides, strong-covalent systems, weak-covalent systems, and metallic bond systems. The inset schematically divides the plane into the \(\sigma\)-bond metallization regime, the DOS-accumulation regime, the weak-coupling region, and the region associated with the competition between lattice stability and strong EPC.
}
\label{Fig_NvsV}
\end{figure*}

Within the present screening framework, the quantity $\Xi$ measures the overall electronic tendency toward strong electron-phonon coupling. However, a large $\Xi$ can arise for two physically distinct reasons: the large scattering strength within two active electronic states via phonon and the number of active states near the Fermi level. To disentangle these two effects, we rewrite $\Xi$ as
\begin{equation}
\Xi^2 = V_{\mathrm{ep}}^{3} N^{*}(\varepsilon_F),
\label{eq:vep_nstar}
\end{equation}
where $V_{\mathrm{ep}}$ is introduced as a measure of the average EPC scattering scale per active electronic state, and
\begin{equation}
N^{*}(\varepsilon_F)=N(\varepsilon_F)/V_{\mathrm{u.c.}}
\end{equation}
is the active density of states per unit volume. This decomposition allows the screened materials to be organized in a more physically transparent way than by $\Xi$ alone. The cubic form of $V_{\mathrm{ep}}$ is adopted only as a convenient dimensional normalization and does not imply any additional physical assumption.

Figure \ref{Fig_NvsV} shows the distribution of the 34 dynamically stable screened candidates with calculated $T_c>10$ K, together with several representative high-pressure hydrides, in the $V_{\mathrm{ep}}-N^{*}(\varepsilon_F)$ plane. The most striking feature is that the compounds do not populate this plane uniformly. Instead, they cluster into two broad regimes. One regime is characterized by large $V_{\mathrm{ep}}$ but only modest $N^{*}(\varepsilon_F)$; the other by large $N^{*}(\varepsilon_F)$ but only moderate $V_{\mathrm{ep}}$. These two regimes correspond naturally to two recurring microscopic routes toward relatively high-$T_c$ phonon-mediated superconductivity. 

The first route is the $\sigma$-bond metallization route, which occupies the upper-left part of Fig. \ref{Fig_NvsV}. The prototypical example is MgB$_2$, whose high $T_c$ is well known to originate from metallized B-B $\sigma$ bonds strongly coupled to in-plane bond-stretching phonons~\cite{MgB2_sig,MgB2_cal,MgB2_arpes}. The same basic tendency is also evident in the boron/carbon-rich candidates and in the high-pressure hydrides included in Fig. \ref{Fig_NvsV}. Although these materials differ greatly in chemistry and structure, they share a common microscopic ingredient: strong directional covalent bonds remain electronically active near the Fermi level. As a result, bond-stretching distortions induce exceptionally large changes in the Hamiltonian, leading to a large average EPC scattering scale $V_{\mathrm{ep}}$. This route is especially favorable for achieving higher $T_c$, because it enhances the intrinsic strength of the electron-lattice interaction itself rather than relying mainly on a large number of available states. 

The second route is the DOS-accumulation route, represented by the lower-right part of Fig. \ref{Fig_NvsV}. Here the average scattering strength is not exceptionally large, but the total EPC can still become substantial because many electronic states participate near $\varepsilon_F$. This regime is common in transition-metal compounds with more metallic bonding or only weakly directional covalent bonding. In such systems, transition-metal $d$ orbitals often generate narrow bands or pronounced peaks in the density of states around the Fermi level. The A15 compounds are representative examples: although their bond-stretching deformation response is less extreme than that in metallized covalent systems, their large Fermi-level DOS amplifies the overall EPC and supports sizable $T_c$ values~\cite{sc_sum,Nb3Ge,Nb3Sn_cal}. This route is therefore an important way to obtain relatively high-$T_c$ superconductivity, although it is generally less effective than the $\sigma$-bond metallization route for pushing $T_c$ to the highest values. 

This landscape also reveals an important constraint. If one were to optimize superconductivity by maximizing both $V_{\mathrm{ep}}$ and $N^{*}(\varepsilon_F)$ simultaneously, the most desirable region would seem to be the upper-right corner of Fig. \ref{Fig_NvsV}. In the present dataset, however, no dynamically stable compounds appear there. This absence suggests that strong EPC and lattice stability are in direct competition. When a system combines very large electron-phonon matrix elements with a high density of electronic states at the Fermi level, the electronic energy gain will be very large when the high-symmetry structure is collapsed, like the formation of charge-density-wave phase or Jahn-Teller-like symmetry lowering~\cite{NbSe2,CVS,CDW}. Such distortions typically reduce $N^{*}(\varepsilon_F)$, relieve the underlying electronic instability, and drive the material away from the region most favorable for superconductivity. This also consists with the substantial attrition observed during the DFPT dynamical-stability verification in Sec. IV.

This observation suggests that relatively high-$T_c$ phonon-mediated superconductors should not be sought by independently maximizing $V_{\mathrm{ep}}$ or $N^{*}(\varepsilon_F)$. A more realistic strategy is to search for compounds located around the stability boundary of Fig. \ref{Fig_NvsV}, where the electron-phonon interaction is already strong but the lattice has not yet become unstable. From this perspective, the materials excluded because of dynamical instability remain scientifically valuable, as pressure, alloying, carrier doping, or chemical substitution may stabilize them while preserving their strong EPC and superconductivity~\cite{presstec}.

\section{Conclusion}

In summary, we developed the static electron-phonon response (SEPR) scheme for screening phonon-mediated superconductors using static orbital Hamiltonians. SEPR can estimate the electronic tendency toward strong electron-phonon coupling (EPC) without explicit phonon perturbation calculations at the initial screening stage. Benchmarks on representative superconductors show that it can capture the main trends and approximate scale of the EPC at a much lower computational cost than traditional DFPT. It therefore provides an efficient and physically transparent tool for high-throughput screening. 

We applied SEPR to more than 36,000 compounds in the MattKeyBond database and then performed full DFPT calculations on the selected materials to verify their dynamical stability and superconducting properties. This workflow identified 34 dynamically stable superconducting candidates with calculated \(T_c > 10\) K. Additionally, these candidates also reveal two recurring routes to relatively high-$T_{c}$ superconductivity: metallized covalent $\sigma$-bonding, which is the more effective route to high $T_{c}$, and Fermi-level density-of-states accumulation, which enhances superconductivity more moderately. These dual routes also highlight a fundamental competition between strong electron-phonon coupling and lattice stability. 

Overall, the SEPR scheme provides both a scalable screening strategy and a unified physical picture of how chemical bonding, electronic structure, and lattice stability shape phonon-mediated superconductivity. More broadly, our results show that static, bond-resolved electronic structures contain sufficient information to guide the search for phonon-mediated superconductors, paving the way for future data-driven materials design.

\begin{acknowledgments}
This work was supported by the Project funded by China Postdoctoral Science Foundation (No. 2022M723355), the Chinese funding sources applied via HPSTAR, the National Natural Science Foundation of China (12488201), and the National Key Research and Development Project of China (2021ZD0301800, 2022YFA1403103).
\end{acknowledgments}

\bibliography{sigma_sc}

@article{MKB,
      title={Bridging Crystal Structure and Material Properties via Bond-Centric Descriptors}, 
      author={Jian-Feng Zhang and Ze-Feng Gao and Xiao-Qi Han and Bo Zhan and Dingshun Lv and Miao Gao and Kai Liu and Xinguo Ren and Zhong-Yi Lu and Tao Xiang},
      year={2026},
      eprint={2603.18876},
      archivePrefix={arXiv},
      primaryClass={cond-mat.mtrl-sci},
      url={https://arxiv.org/abs/2603.18876}, 
      journal = {arXiv},
}

@article{pwscf,
	doi = {10.1088/0953-8984/21/39/395502},
	url = {https://doi.org/10.1088/0953-8984/21/39/395502},
	year = 2009,
	month = {sep},
	publisher = {{IOP} Publishing},
	volume = {21},
	number = {39},
	pages = {395502},
	author = {Paolo Giannozzi and Stefano Baroni and Nicola Bonini and Matteo Calandra and Roberto Car and Carlo Cavazzoni and Davide Ceresoli and Guido L Chiarotti and Matteo Cococcioni and Ismaila Dabo and Andrea Dal Corso and Stefano de Gironcoli and Stefano Fabris and Guido Fratesi and Ralph Gebauer and Uwe Gerstmann and Christos Gougoussis and vAnton Kokalj and Michele Lazzeri and Layla Martin-Samos and Nicola Marzari and Francesco Mauri and Riccardo Mazzarello and Stefano Paolini and Alfredo Pasquarello and Lorenzo Paulatto and Carlo Sbraccia and Sandro Scandolo and Gabriele Sclauzero and Ari P Seitsonen and Alexander Smogunov and Paolo Umari and Renata M Wentzcovitch},
	title = {{QUANTUM} {ESPRESSO}: a modular and open-source software project for quantum simulations of materials},
	journal = {J. Phys.: Condens. Matter},
}

@article{epw,
title = {\text{EPW}: A program for calculating the electron–phonon coupling using maximally localized Wannier functions},
journal = {Comput. Phys. Commun.},
volume = {181},
number = {12},
pages = {2140-2148},
year = {2010},
issn = {0010-4655},
doi = {https://doi.org/10.1016/j.cpc.2010.08.027},
url = {https://www.sciencedirect.com/science/article/pii/S0010465510003218},
author = {Jesse Noffsinger and Feliciano Giustino and Brad D. Malone and Cheol-Hwan Park and Steven G. Louie and Marvin L. Cohen},

}

@article{bcs,
  title = {Microscopic Theory of Superconductivity},
  author = {Bardeen, J. and Cooper, L. N. and Schrieffer, J. R.},
  journal = {Phys. Rev.},
  volume = {106},
  issue = {1},
  pages = {162--164},
  numpages = {0},
  year = {1957},
  month = {Apr},
  publisher = {American Physical Society},
  doi = {10.1103/PhysRev.106.162},
  url = {https://link.aps.org/doi/10.1103/PhysRev.106.162}
}

@article{cup1,
  title = {Correlated electrons in high-temperature superconductors},
  author = {Dagotto, Elbio},
  journal = {Rev. Mod. Phys.},
  volume = {66},
  issue = {3},
  pages = {763--840},
  numpages = {0},
  year = {1994},
  month = {Jul},
  publisher = {American Physical Society},
  doi = {10.1103/RevModPhys.66.763},
  url = {https://link.aps.org/doi/10.1103/RevModPhys.66.763}
}

@article{cup2,
  title = {Doping a Mott insulator: Physics of high-temperature superconductivity},
  author = {Lee, Patrick A. and Nagaosa, Naoto and Wen, Xiao-Gang},
  journal = {Rev. Mod. Phys.},
  volume = {78},
  issue = {1},
  pages = {17--85},
  numpages = {0},
  year = {2006},
  month = {Jan},
  publisher = {American Physical Society},
  doi = {10.1103/RevModPhys.78.17},
  url = {https://link.aps.org/doi/10.1103/RevModPhys.78.17}
}

@article{iro1,
author = {Kamihara, Yoichi and Watanabe, Takumi and Hirano, Masahiro and Hosono, Hideo},
title = {Iron-Based Layered Superconductor $\text {LaO}_{1-x}\text{F}_x\text{FeAs}$ ($x$ = 0.05-0.12) with \text{T}$_c$ = 26 \text{K}},
journal = {J. Am. Chem. Soc.},
volume = {130},
number = {11},
pages = {3296-3297},
year = {2008},
doi = {10.1021/ja800073m},
URL = {https://doi.org/10.1021/ja800073m}
}

@article{iro2,
  title = {Arsenic-bridged antiferromagnetic superexchange interactions in LaFeAsO},
  author = {Ma, Fengjie and Lu, Zhong-Yi and Xiang, Tao},
  journal = {Phys. Rev. B},
  volume = {78},
  issue = {22},
  pages = {224517},
  numpages = {6},
  year = {2008},
  month = {Dec},
  publisher = {American Physical Society},
  doi = {10.1103/PhysRevB.78.224517},
  url = {https://link.aps.org/doi/10.1103/PhysRevB.78.224517}
}

@article{iro3,
author = {Fong-Chi Hsu  and Jiu-Yong Luo  and Kuo-Wei Yeh  and Ta-Kun Chen  and Tzu-Wen Huang  and Phillip M. Wu  and Yong-Chi Lee  and Yi-Lin Huang  and Yan-Yi Chu  and Der-Chung Yan  and Maw-Kuen Wu },
title = {Superconductivity in the \text{PbO}-type structure $\alpha$-\text{FeSe}},
journal = {Proc. Natl. Acad. Sci.},
volume = {105},
number = {38},
pages = {14262-14264},
year = {2008},
doi = {10.1073/pnas.0807325105},
URL = {https://www.pnas.org/doi/abs/10.1073/pnas.0807325105},
}

@article{Nick,
  title = {{Signatures of superconductivity near 80 K in a nickelate under high pressure}},
  author={Hualei Sun and Mengwu Huo and Xunwu Hu and Jingyuan Li and Zengjia Liu and Yifeng Han and Lingyun Tang and Zhongquan Mao and Pengtao Yang and Bosen Wang and Jinguang Cheng and Dao-Xin Yao and Guang-Ming Zhang and Meng Wang},
  journal={Nature},
  year={2023},
  volume={621},
  pages={493 - 498},
  url={https://api.semanticscholar.org/CorpusID:259843168}
}

@article{metalHpre,
  title = {Metallic Hydrogen: A High-Temperature Superconductor?},
  author = {Ashcroft, N. W.},
  journal = {Phys. Rev. Lett.},
  volume = {21},
  issue = {26},
  pages = {1748--1749},
  numpages = {0},
  year = {1968},
  month = {Dec},
  publisher = {American Physical Society},
  doi = {10.1103/PhysRevLett.21.1748},
  url = {https://link.aps.org/doi/10.1103/PhysRevLett.21.1748}
}

@Article{presstec,
author={Dubrovinsky, Leonid
and Dubrovinskaia, Natalia
and Prakapenka, Vitali B.
and Abakumov, Artem M.},
title={Implementation of micro-ball nanodiamond anvils for high-pressure studies above 6 \text{Mbar}},
journal={Nat. Commun.},
year={2012},
month={Oct},
day={23},
volume={3},
number={1},
pages={1163},
issn={2041-1723},
doi={10.1038/ncomms2160},
url={https://doi.org/10.1038/ncomms2160}
}

@Article{MgB2_1st,
author={Nagamatsu, Jun
and Nakagawa, Norimasa
and Muranaka, Takahiro
and Zenitani, Yuji
and Akimitsu, Jun},
title={Superconductivity at 39 \text{K} in magnesium diboride},
journal={Nature},
year={2001},
month={Mar},
day={01},
volume={410},
number={6824},
pages={63-64},
issn={1476-4687},
doi={10.1038/35065039},
url={https://doi.org/10.1038/35065039}
}

@Article{SH3_1st,
author={Drozdov, A. P.
and Eremets, M. I.
and Troyan, I. A.
and Ksenofontov, V.
and Shylin, S. I.},
title={Conventional superconductivity at 203 kelvin at high pressures in the sulfur hydride system},
journal={Nature},
year={2015},
month={Sep},
day={01},
volume={525},
number={7567},
pages={73-76},
issn={1476-4687},
doi={10.1038/nature14964},
url={https://doi.org/10.1038/nature14964}
}

@article{maprl,
  title = {Hydrogen Clathrate Structures in Rare Earth Hydrides at High Pressures: Possible Route to Room-Temperature Superconductivity},
  author = {Peng, Feng and Sun, Ying and Pickard, Chris J. and Needs, Richard J. and Wu, Qiang and Ma, Yanming},
  journal = {Phys. Rev. Lett.},
  volume = {119},
  issue = {10},
  pages = {107001},
  numpages = {6},
  year = {2017},
  month = {Sep},
  publisher = {American Physical Society},
  doi = {10.1103/PhysRevLett.119.107001},
  url = {https://link.aps.org/doi/10.1103/PhysRevLett.119.107001}
}

@article{maprl2,
  title = {Route to a Superconducting Phase above Room Temperature in Electron-Doped Hydride Compounds under High Pressure},
  author = {Sun, Ying and Lv, Jian and Xie, Yu and Liu, Hanyu and Ma, Yanming},
  journal = {Phys. Rev. Lett.},
  volume = {123},
  issue = {9},
  pages = {097001},
  numpages = {5},
  year = {2019},
  month = {Aug},
  publisher = {American Physical Society},
  doi = {10.1103/PhysRevLett.123.097001},
  url = {https://link.aps.org/doi/10.1103/PhysRevLett.123.097001}
}

@article{LaH10_1st,
  title = {Evidence for Superconductivity above 260 \text{K} in \text{Lanthanum Superhydride at Megabar Pressures}},
  author = {Somayazulu, Maddury and Ahart, Muhtar and Mishra, Ajay K. and Geballe, Zachary M. and Baldini, Maria and Meng, Yue and Struzhkin, Viktor V. and Hemley, Russell J.},
  journal = {Phys. Rev. Lett.},
  volume = {122},
  issue = {2},
  pages = {027001},
  numpages = {6},
  year = {2019},
  month = {Jan},
  publisher = {American Physical Society},
  doi = {10.1103/PhysRevLett.122.027001},
  url = {https://link.aps.org/doi/10.1103/PhysRevLett.122.027001}
}

@Article{LaH10_2nd,
author={Drozdov, A. P.
and Kong, P. P.
and Minkov, V. S.
and Besedin, S. P.
and Kuzovnikov, M. A.
and Mozaffari, S.
and Balicas, L.
and Balakirev, F. F.
and Graf, D. E.
and Prakapenka, V. B.
and Greenberg, E.
and Knyazev, D. A.
and Tkacz, M.
and Eremets, M. I.},
title={Superconductivity at 250 \text{K} in lanthanum hydride under high pressures},
journal={Nature},
year={2019},
month={May},
day={01},
volume={569},
number={7757},
pages={528-531},
issn={1476-4687},
doi={10.1038/s41586-019-1201-8},
url={https://doi.org/10.1038/s41586-019-1201-8}
}

@Article{YH9_1st,
author={Kong, Panpan
and Minkov, Vasily S.
and Kuzovnikov, Mikhail A.
and Drozdov, Alexander P.
and Besedin, Stanislav P.
and Mozaffari, Shirin
and Balicas, Luis
and Balakirev, Fedor Fedorovich
and Prakapenka, Vitali B.
and Chariton, Stella
and Knyazev, Dmitry A.
and Greenberg, Eran
and Eremets, Mikhail I.},
title={Superconductivity up to 243 \text{K} in the yttrium-hydrogen system under high pressure},
journal={Nat. Commun.},
year={2021},
month={Aug},
day={20},
volume={12},
number={1},
pages={5075},
issn={2041-1723},
doi={10.1038/s41467-021-25372-2},
url={https://doi.org/10.1038/s41467-021-25372-2}
}

@article{CaH6_1st,
  title = {High-Temperature Superconducting Phase in Clathrate Calcium Hydride $\text{CaH}_6$ up to 215 \text{K} at a Pressure of 172 \text{GPa}},
  author = {Ma, Liang and Wang, Kui and Xie, Yu and Yang, Xin and Wang, Yingying and Zhou, Mi and Liu, Hanyu and Yu, Xiaohui and Zhao, Yongsheng and Wang, Hongbo and Liu, Guangtao and Ma, Yanming},
  journal = {Phys. Rev. Lett.},
  volume = {128},
  issue = {16},
  pages = {167001},
  numpages = {6},
  year = {2022},
  month = {Apr},
  publisher = {American Physical Society},
  doi = {10.1103/PhysRevLett.128.167001},
  url = {https://link.aps.org/doi/10.1103/PhysRevLett.128.167001}
}

@Article{CaH6_2nd,
author={Li, Zhiwen
and He, Xin
and Zhang, Changling
and Wang, Xiancheng
and Zhang, Sijia
and Jia, Yating
and Feng, Shaomin
and Lu, Ke
and Zhao, Jianfa
and Zhang, Jun
and Min, Baosen
and Long, Youwen
and Yu, Richeng
and Wang, Luhong
and Ye, Meiyan
and Zhang, Zhanshuo
and Prakapenka, Vitali
and Chariton, Stella
and Ginsberg, Paul A.
and Bass, Jay
and Yuan, Shuhua
and Liu, Haozhe
and Jin, Changqing},
title={Superconductivity above 200 \text{K} discovered in superhydrides of calcium},
journal={Nat. Commun.},
year={2022},
month={May},
day={23},
volume={13},
number={1},
pages={2863},
issn={2041-1723},
doi={10.1038/s41467-022-30454-w},
url={https://doi.org/10.1038/s41467-022-30454-w}
}

@article{dfptreview,
  title = {Electron-phonon interactions from first principles},
  author = {Giustino, Feliciano},
  journal = {Rev. Mod. Phys.},
  volume = {89},
  issue = {1},
  pages = {015003},
  numpages = {63},
  year = {2017},
  month = {Feb},
  publisher = {American Physical Society},
  doi = {10.1103/RevModPhys.89.015003},
  url = {https://link.aps.org/doi/10.1103/RevModPhys.89.015003}
}

@article{dfptreview2,
  title = {Phonons and related crystal properties from density-functional perturbation theory},
  author = {Baroni, Stefano and de Gironcoli, Stefano and Dal Corso, Andrea and Giannozzi, Paolo},
  journal = {Rev. Mod. Phys.},
  volume = {73},
  issue = {2},
  pages = {515--562},
  numpages = {0},
  year = {2001},
  month = {Jul},
  publisher = {American Physical Society},
  doi = {10.1103/RevModPhys.73.515},
  url = {https://link.aps.org/doi/10.1103/RevModPhys.73.515}
}

@Article{ML_sc1,
author={Stanev, Valentin
and Oses, Corey
and Kusne, A. Gilad
and Rodriguez, Efrain
and Paglione, Johnpierre
and Curtarolo, Stefano
and Takeuchi, Ichiro},
title={Machine learning modeling of superconducting critical temperature},
journal={npj Computational Materials},
year={2018},
month={Jun},
day={28},
volume={4},
number={1},
pages={29},
issn={2057-3960},
doi={10.1038/s41524-018-0085-8},
url={https://doi.org/10.1038/s41524-018-0085-8}
}

@Article{ML_sc2,
author={Xie, S. R.
and Quan, Y.
and Hire, A. C.
and Deng, B.
and DeStefano, J. M.
and Salinas, I.
and Shah, U. S.
and Fanfarillo, L.
and Lim, J.
and Kim, J.
and Stewart, G. R.
and Hamlin, J. J.
and Hirschfeld, P. J.
and Hennig, R. G.},
title={Machine learning of superconducting critical temperature from Eliashberg theory},
journal={npj Computational Materials},
year={2022},
month={Jan},
day={25},
volume={8},
number={1},
pages={14},
issn={2057-3960},
doi={10.1038/s41524-021-00666-7},
url={https://doi.org/10.1038/s41524-021-00666-7}
}

@article{dft1,
  title = {Inhomogeneous Electron Gas},
  author = {Hohenberg, P. and Kohn, W.},
  journal = {Phys. Rev.},
  volume = {136},
  issue = {3B},
  pages = {B864--B871},
  numpages = {0},
  year = {1964},
  month = {Nov},
  publisher = {American Physical Society},
  doi = {10.1103/PhysRev.136.B864},
  url = {https://link.aps.org/doi/10.1103/PhysRev.136.B864}
}

@article{dft2,
  title = {Self-Consistent Equations Including Exchange and Correlation Effects},
  author = {Kohn, W. and Sham, L. J.},
  journal = {Phys. Rev.},
  volume = {140},
  issue = {4A},
  pages = {A1133--A1138},
  numpages = {0},
  year = {1965},
  month = {Nov},
  publisher = {American Physical Society},
  doi = {10.1103/PhysRev.140.A1133},
  url = {https://link.aps.org/doi/10.1103/PhysRev.140.A1133}
}

@article{mcmillan1,
  title = {Transition Temperature of Strong-Coupled Superconductors},
  author = {McMillan, W. L.},
  journal = {Phys. Rev.},
  volume = {167},
  issue = {2},
  pages = {331--344},
  numpages = {0},
  year = {1968},
  month = {Mar},
  publisher = {American Physical Society},
  doi = {10.1103/PhysRev.167.331},
  url = {https://link.aps.org/doi/10.1103/PhysRev.167.331}
}

@article{mcmillan2,
  title = {Transition temperature of strong-coupled superconductors reanalyzed},
  author = {Allen, P. B. and Dynes, R. C.},
  journal = {Phys. Rev. B},
  volume = {12},
  issue = {3},
  pages = {905--922},
  numpages = {0},
  year = {1975},
  month = {Aug},
  publisher = {American Physical Society},
  doi = {10.1103/PhysRevB.12.905},
  url = {https://link.aps.org/doi/10.1103/PhysRevB.12.905}
}

@article{NbTi,
  title = {First-principles study of the robust superconducting state of NbTi alloys under ultrahigh pressures},
  author = {Zhang, Jian-Feng and Gao, Miao and Liu, Kai and Lu, Zhong-Yi},
  journal = {Phys. Rev. B},
  volume = {102},
  issue = {19},
  pages = {195140},
  numpages = {6},
  year = {2020},
  month = {Nov},
  publisher = {American Physical Society},
  doi = {10.1103/PhysRevB.102.195140},
  url = {https://link.aps.org/doi/10.1103/PhysRevB.102.195140}
}

@article{Nb_alloy2,
author = {Guo, Jing and Lin, Gongchang and Cai, Shu and Xi, Chuanying and Zhang, Changjin and Sun, Wanshuo and Wang, Qiuliang and Yang, Ke and Li, Aiguo and Wu, Qi and Zhang, Yuheng and Xiang, Tao and Cava, Robert Joseph and Sun, Liling},
title = {Record-High Superconductivity in Niobium–Titanium Alloy},
journal = {Advanced Materials},
volume = {31},
number = {11},
pages = {1807240},
doi = {https://doi.org/10.1002/adma.201807240},
url = {https://advanced.onlinelibrary.wiley.com/doi/abs/10.1002/adma.201807240},
year = {2019}
}

@article{Nb_alloy1,
author = {Jing Guo  and Honghong Wang  and Fabian von Rohr  and Zhe Wang  and Shu Cai  and Yazhou Zhou  and Ke Yang  and Aiguo Li  and Sheng Jiang  and Qi Wu  and Robert J. Cava  and Liling Sun },
title = {Robust zero resistance in a superconducting high-entropy alloy at pressures up to 190 GPa},
journal = {Proceedings of the National Academy of Sciences},
volume = {114},
number = {50},
pages = {13144-13147},
year = {2017},
doi = {10.1073/pnas.1716981114},
URL = {https://www.pnas.org/doi/abs/10.1073/pnas.1716981114}}

@article{cwf,
  title = {Closest Wannier functions to a given set of localized orbitals},
  author = {Ozaki, Taisuke},
  journal = {Phys. Rev. B},
  volume = {110},
  issue = {12},
  pages = {125115},
  numpages = {13},
  year = {2024},
  month = {Sep},
  publisher = {American Physical Society},
  doi = {10.1103/PhysRevB.110.125115},
  url = {https://link.aps.org/doi/10.1103/PhysRevB.110.125115}
}

@article{NP_topo,
  title={Non-trivial quantum geometry and the strength of electron--phonon coupling},
  author={Yu, Jiabin and Ciccarino, Christopher J and Bianco, Raffaello and Errea, Ion and Narang, Prineha and Bernevig, B Andrei},
  journal={Nature Physics},
  volume={20},
  number={8},
  pages={1262--1268},
  year={2024},
  publisher={Nature Publishing Group UK London}
}

@article{bkbo_pre,
  title = {Superconducting-insulating phase transition in pressurized ${\mathrm{Ba}}_{1\ensuremath{-}x}{\mathrm{K}}_{x}{\mathrm{BiO}}_{3}$},
  author = {Han, Jinyu and Zhu, Xiangde and Zhang, Jianfeng and Cai, Shu and Wang, Luhong and Gao, Yang and Liu, Fuyang and Liu, Haozhe and Kawaguchi, Saori I. and Guo, Jing and Zhou, Yazhou and Zhao, Jinyu and Wang, Pengyu and Cao, Lixin and Tian, Mingliang and Wu, Qi and Xiang, Tao and Sun, Liling},
  journal = {Phys. Rev. B},
  volume = {111},
  issue = {2},
  pages = {L020509},
  numpages = {6},
  year = {2025},
  month = {Jan},
  publisher = {American Physical Society},
  doi = {10.1103/PhysRevB.111.L020509},
  url = {https://link.aps.org/doi/10.1103/PhysRevB.111.L020509}
}

@Article{icsd,
author={Zagorac, D.
        and Müller, H.
        and Ruehl, S.
        and Zagorac, J. 
        and Rehme, S.},
title={Recent developments in the Inorganic Crystal Structure Database: theoretical crystal structure data and related features},
journal={J. Appl. Cryst.},
year={2019},
volume={52},
pages={918-925},
doi={10.1107/S160057671900997X},
}

@article{mp,
    author = {Jain, Anubhav and Ong, Shyue Ping and Hautier, Geoffroy and Chen, Wei and Richards, William Davidson and Dacek, Stephen and Cholia, Shreyas and Gunter, Dan and Skinner, David and Ceder, Gerbrand and Persson, Kristin A.},
    title = {Commentary: The Materials Project: A materials genome approach to accelerating materials innovation},
    journal = {APL Materials},
    volume = {1},
    number = {1},
    pages = {011002},
    year = {2013},
    month = {07},
    issn = {2166-532X},
    doi = {10.1063/1.4812323},
    url = {https://doi.org/10.1063/1.4812323},
}

@article{Nb3Ge,
  title={Superconductivity in Nb-Ge films above 22 K},
  author={Gavaler, JR},
  journal={Applied Physics Letters},
  volume={23},
  number={8},
  pages={480--482},
  year={1973}
}

@article{sc_sum,
  title={Survey of superconductive materials and critical evaluation of selected properties},
  author={Roberts, Benjamin Washington},
  journal={Journal of Physical and Chemical Reference Data},
  volume={5},
  number={3},
  pages={581--822},
  year={1976},
  publisher={American Institute of Physics for the National Institute of Standards and~…}
}

@Article{bkbo,
author={Cava, R. J.
and Batlogg, B.
and Krajewski, J. J.
and Farrow, R.
and Rupp, L. W.
and White, A. E.
and Short, K.
and Peck, W. F.
and Kometani, T.},
title={Superconductivity near 30 K without copper: the Ba0.6K0.4BiO3 perovskite},
journal={Nature},
year={1988},
month={Apr},
day={01},
volume={332},
number={6167},
pages={814-816},
issn={1476-4687},
doi={10.1038/332814a0},
url={https://doi.org/10.1038/332814a0}
}

@article{PBE,
  title = {Generalized Gradient Approximation Made Simple},
  author = {Perdew, John P. and Burke, Kieron and Ernzerhof, Matthias},
  journal = {Phys. Rev. Lett.},
  volume = {77},
  issue = {18},
  pages = {3865--3868},
  numpages = {0},
  year = {1996},
  month = {Oct},
  publisher = {American Physical Society},
  doi = {10.1103/PhysRevLett.77.3865},
  url = {https://link.aps.org/doi/10.1103/PhysRevLett.77.3865}
}

@article{bkbo_cal1,
  title={Correlation-enhanced electron-phonon coupling: Applications of GW and screened hybrid functional to bismuthates, chloronitrides, and other high-T c superconductors},
  author={Yin, ZP and Kutepov, A and Kotliar, G},
  journal={Physical Review X},
  volume={3},
  number={2},
  pages={021011},
  year={2013},
  publisher={APS}
}

@article{bkbo_cal2,
  title={Electron-phonon coupling from ab initio linear-response theory within the GW method: Correlation-enhanced interactions and superconductivity in Ba 1- x K x BiO 3},
  author={Li, Zhenglu and Antonius, Gabriel and Wu, Meng and Da Jornada, Felipe H and Louie, Steven G},
  journal={Physical review letters},
  volume={122},
  number={18},
  pages={186402},
  year={2019},
  publisher={APS}
}

@article{MgB2_sig,
  title={Superconductivity of MgB 2: covalent bonds driven metallic},
  author={An, JM and Pickett, WE},
  journal={Physical Review Letters},
  volume={86},
  number={19},
  pages={4366},
  year={2001},
  publisher={APS}
}

@article{MgB2_cal,
  title={The origin of the anomalous superconducting properties of MgB2},
  author={Choi, Hyoung Joon and Roundy, David and Sun, Hong and Cohen, Marvin L and Louie, Steven G},
  journal={Nature},
  volume={418},
  number={6899},
  pages={758--760},
  year={2002},
  publisher={Nature Publishing Group UK London}
}

@article{MgB2_arpes,
  title={The origin of multiple superconducting gaps in MgB2},
  author={Souma, S and Machida, Y and Sato, T and Takahashi, T and Matsui, H and Wang, S-C and Ding, H and Kaminski, A and Campuzano, JC and Sasaki, S and others},
  journal={Nature},
  volume={423},
  number={6935},
  pages={65--67},
  year={2003},
  publisher={Nature Publishing Group UK London}
}

@article{Nb3Sn_cal,
  title={An ab Initio Answer to Long-Debated Questions about Superconducting Nb 3 Sn},
  author={Cucciari, Alessio and Boeri, Lilia},
  journal={PRX Energy},
  volume={5},
  number={1},
  pages={013006},
  year={2026},
  publisher={APS}
}

@article{Nb_cal,
  title={Electron-phonon interactions and related physical properties of metals from linear-response theory},
  author={Savrasov, S Yu and Savrasov, D Yu},
  journal={Physical Review B},
  volume={54},
  number={23},
  pages={16487},
  year={1996},
  publisher={APS}
}

@article{eliashberg,
  title={Anisotropic migdal-eliashberg theory using wannier functions},
  author={Margine, Elena Roxana and Giustino, Feliciano},
  journal={Physical Review B—Condensed Matter and Materials Physics},
  volume={87},
  number={2},
  pages={024505},
  year={2013},
  publisher={APS}
}

@article{CVS,
  title={First-principles study of the double-dome superconductivity in the kagome material CsV 3 Sb 5 under pressure},
  author={Zhang, Jian-Feng and Liu, Kai and Lu, Zhong-Yi},
  journal={Physical Review B},
  volume={104},
  number={19},
  pages={195130},
  year={2021},
  publisher={APS}
}

@article{CDW,
  title={Fermi surface nesting and the origin of charge density waves in metals},
  author={Johannes, MD and Mazin, II},
  journal={Physical Review B—Condensed Matter and Materials Physics},
  volume={77},
  number={16},
  pages={165135},
  year={2008},
  publisher={APS}
}

@article{NbSe2,
  title={Extended phonon collapse and the origin of the charge-density wave in 2 H-NbSe 2},
  author={Weber, F and Rosenkranz, S and Castellan, J-P and Osborn, R and Hott, R and Heid, R and Bohnen, K-P and Egami, T and Said, AH and Reznik, D},
  journal={Physical review letters},
  volume={107},
  number={10},
  pages={107403},
  year={2011},
  publisher={APS}
}

@article{A15,
  title={Superconductivity in the A15 structure},
  author={Stewart, Greygory R},
  journal={Physica C: Superconductivity and its Applications},
  volume={514},
  pages={28--35},
  year={2015},
  publisher={Elsevier}
}

@Article{deepH,
author={Li, He
and Wang, Zun
and Zou, Nianlong
and Ye, Meng
and Xu, Runzhang
and Gong, Xiaoxun
and Duan, Wenhui
and Xu, Yong},
title={Deep-learning density functional theory Hamiltonian for efficient ab initio electronic-structure calculation},
journal={Nature Computational Science},
year={2022},
month={Jun},
day={01},
volume={2},
number={6},
pages={367-377},
issn={2662-8457},
doi={10.1038/s43588-022-00265-6},
url={https://doi.org/10.1038/s43588-022-00265-6}
}

@misc{sigmados,
      title={AI-accelerated metallized $\sigma$-bonding screening for superconductor discovery}, 
      author={Zechen Tang and Wen-Han Dong and Baochun Wu and Jian-Feng Zhang and Yuxiang Wang and Yang Li and Honggeng Tao and Qiyu Zeng and Chong Wang and Chen Si and Zhong-Yi Lu and Wenhui Duan and Tao Xiang and Yong Xu},
      year={2026},
      eprint={2606.21251},
      archivePrefix={arXiv},
      primaryClass={physics.comp-ph},
      url={https://arxiv.org/abs/2606.21251}, 
}

@article{MoB2,
title = {The record-high robust superconducting temperature of 32 K in $\alpha$-MoB2 under high pressure},
journal = {Materials Today},
volume = {97},
pages = {103386},
year = {2026},
issn = {1369-7021},
doi = {https://doi.org/10.1016/j.mattod.2026.103386},
url = {https://www.sciencedirect.com/science/article/pii/S1369702126002324},
author = {Xuqiang Liu and Cuiying Pei and Jianfeng Zhang and Qi Wang and Nana Li and Limin Yan and Jiayi Guan and Yiming Wang and Mingtao Li and Haoming Cheng and Hechang Lei and Kai Liu and Yanpeng Qi and Wenge Yang},

}

@article{RSAVS,
  title = {RSAVS superconductors: Materials with a superconducting state that is robust against large volume shrinkage},
  author = {Huang, Cheng and Guo, Jing and Zhang, Jianfeng and Stolze, Karoline and Cai, Shu and Liu, Kai and Weng, Hongming and Lu, Zhongyi and Wu, Qi and Xiang, Tao and Cava, Robert J. and Sun, Liling},
  journal = {Phys. Rev. Mater.},
  volume = {4},
  issue = {7},
  pages = {071801(R)},
  numpages = {5},
  year = {2020},
  month = {Jul},
  publisher = {American Physical Society},
  doi = {10.1103/PhysRevMaterials.4.071801},
  url = {https://link.aps.org/doi/10.1103/PhysRevMaterials.4.071801}
}

\end{document}